\documentclass[]{aa}
\usepackage{natbib}
\usepackage{psfig}
\usepackage{txfonts}
\def\EBV{\mbox{E$_{\rm BV}$}}
\def\EBVl{\mbox{E$^\infty_{\rm BV}$}}

\def\AV{\mbox{A$_{\rm V}$}}

\def\HH{\mbox{H$_2$}}

\def\nH2{\mbox{${\rm n}_\HH}$}

\def\NH2{{\rm N}({\rm H}_2)}
\def\pccc{~{\rm cm}^{-3}} 
 
\def\pcc {~{\rm cm}^{-2}}

\def\Tstar {\mbox{${\rm T}_{\rm r}^*$}}
\def\Tsub#1 {\mbox{${\rm T}_{\rm #1}$}}
\def\TK  {\Tsub K }
\def\TB  {\Tsub B }

\def\Texc {\Tsub ex }

\def\arcsec{\mbox{$^{\prime\prime}$}} \def\arcmin{\mbox{$^{\prime}$}}

\def\degr{$^{\rm o}$}
\def\p{\mbox{$^+$}}
\def\cotw {\mbox{$^{12}$CO}}
\def\coth {\mbox{$^{13}$CO}}

\def\WCO{\mbox{W$_{\rm CO}$}}
\def\ACO{\mbox{A$_{\rm CO}$}}

\def\h13cop{\mbox{{H$^{13}$CO\p}}}

\def\c3h2{\mbox{C$_3$H$_2$}}
 
 \def\R0{R$_0$} 
  
  \def\deg{{}^\circ}
\def\ddeg{{}^\circ\kern-.1em}

\def\kms{\mbox{km\,s$^{-1}$}}

\def\E#1 {$10^{#1}$}
\def\E#1 {E{#1}}
\def\P#1,{$\nH2\TK~=~#1\times~10^4\pccc$~K}
\def\ec#1,#2,#3,{#1\,(#2)\E{#3}}

\def\zoph{$\zeta$ Oph}

\def\H3{\mbox{H$_3$}}

\def\RH2{\mbox{R$_{\rm G}$}}
\def\fH2{\mbox{f$_{\HH}$}}
\def\FH2{\mbox{F$_{\HH}$}}

\def\g13{\mbox{g$_{13}$}} 

\begin{document}
\sloppy
\title{A CO J=1-0 Survey of common optical/$uv$ absorption sightlines
  \thanks{Based on observations obtained with the ARO Kitt Peak 12m 
   telescope.}} 

\author{H. S. Liszt\inst{1} }
\institute{National Radio Astronomy Observatory,
           520 Edgemont Road,
           Charlottesville, VA,
           USA 22903-2475}


\date{received \today}
\offprints{H. S. Liszt}
\mail{hliszt@nrao.edu}
%
\abstract
   { Comparison of optical/uv absorption line data with high-resolution profiles
  of mm-wave CO emission provides complementary information on the
  absorbing gas, as toward \zoph.  Over the past thirty years a wealth of observations 
 of CO and other molecules in optical/uv absorption in 
  diffuse clouds has accumulated  for which no 
  comparable CO emission line data exist.}
{To acquire mm-wave J=1-0 CO emission line profiles toward a substantial sample 
  of commonly-studied optical/uv absorption line targets and to compare 
 with the properties of the absorbing gas, especially
 the  predicted emission line strengths.}
 {Using the ARO 12m telescope, we observed mm-wavelength J=1-0 CO emission 
  with spectral resolution R $\approx 3\times 10^6$ and spatial resolution 
   1\arcmin\ toward a sample of 110 lines of sight previously studied in 
  optical/uv absorption lines of CO, \HH, CH, $etc.$}
 {Interstellar CO emission was detected along 65 of the 110 lines of
 sight surveyed and there is a general superabundance of CO emission 
  given the distribution of galactic latitudes in the survey sample.
  Much of the  emission is optically thick or very intense
 and must emanate from dark clouds or warm dense gas near HII regions.}   
{ Judging from the statistical superabundance of CO emission, 
   seen also in the total line of sight reddening,
  the OB star optical/$uv$ absorption line targets must be
  physically associated with the large quantities of neutral
  gas whose CO emission was detected, in which case they are probably influencing
  the absorbing gas by heating and/or photoionizing it.  
  This explains why  CO/\HH\ and \cotw/\coth\
  ratios differ somewhat between $uv$ and mm-wave absorption line 
  studies.  Because the lines of sight have been preselected to have 
  \AV\ $\la$ 1 mag, relatively little of the associated material 
  actually occults the targets, making it difficult for CO emission
 line observations to isolate the foreground gas contribution.}
\keywords{ interstellar medium -- molecules }

\maketitle

\section {Introduction.}

The presence of gas-phase carbon monoxide (CO) in diffuse clouds
 at \AV\ $\la$ 1 mag was shown by \cite{SmiSte71} shortly after the 
discovery of interstellar CO itself $via$ the intense mm-wave CO 
J=1-0 rotational emission in the Orion Nebula \citep{WilJef+70}.
Although the fractional abundance of CO is relatively small in
diffuse clouds, typically $10^{-7} < $ N(CO)/N(\HH) $ < 10^{-5}$
so that N(CO) $<<$ N(C\p) and at most a few percent of the free 
gas-phase carbon is actually in CO, the presence of even this much 
CO has proved a consistent challenge to interstellar chemistry 
\citep{GlaLan76,BlaDal77,BalLan82,GlaHug+85,vDiBla88,LeeHer+96}.

It is only quite recently that substantial numbers of sightlines have 
been studied with high sensitivity in $uv$ absorption with HST and FUSE
\citep{SonWel+07,BurFra+07,SheRog+07,SheRog+08} and in mm-wave absorption 
against extragalactic continuum sources \citep{LisLuc98} at the PdBI
so that the formation, excitation and fractionation of CO in diffuse 
clouds can be considered in a systematic fashion 
($ibid$ and \cite{Lis07CO}); compare with the much earlier survey 
of \cite{FedGla+80}.  Notwithstanding the 
long interval since its discovery, carbon monoxide is now the molecule 
whose chemistry and abundance relative to \HH\ are the most exhaustively
studied in the diffuse interstellar medium (ISM).  With much scatter, 
N(CO) $\propto$ N(\HH)$^2$ for $10^{12}\pcc\ < $ N(CO) $ 
3\times 10^{16}\pcc$, $10^{19} <$ N(\HH) $\la 2\times 10^{21}\pcc$.

From the CO column densities and rotational excitation temperatures 
observed in  absorption, it is straightforward to calculate that the 
integrated brightness temperature of CO J=1-0 rotational emission in 
diffuse gas should vary as 
\WCO\ = $\int\TB dv \approx $ 1 K \kms\ $ \times {\rm N(CO)}/10^{15}\pcc$
for N(CO) $\la 10^{16}\pcc$ and \WCO\ as strong as 10-12 K \kms\
\citep{Lis07CO}.  This was actually observed when comparing CO emission 
and 
absorption at mm-wavelengths toward distant extragalactic background 
sources, where separating foreground and background gas is not an issue
\citep{LisLuc98}.  It is also correct along the archetypal line
of sight toward \zoph\ at N(CO) $= 2.2\times 10^{15}~\pcc$.
Proportionality between \WCO\ and N(CO) is a very general consequence 
of weak excitation, {\it i.e.} \Texc\ $<<$ \TK\ \citep{GolKwa74} but the
normalization is determined by ambient physical conditions.
As discussed below (see Sect 5), \WCO/N(CO) is about 50 times 
higher in diffuse clouds than toward Orion A or typical dark clouds.

The CO emission from absorbing foreground material in diffuse clouds 
should be readily detectable if N(CO) $\ga  3\times 10^{14}\pcc$ and 
rather strong, comparable to the emission from dark clouds, 
whenever N(CO) $\ga 10^{15} \pcc$.  Morever, if CO emission is
observable, the high spectral resolution available in mm-wave 
spectra, typically R $\approx 3\times10^6$, provides
complementary information to that available in lower-resolution 
absorption spectra \citep{KnaJur76,Lis79,WanPen+82,LanGla+87,
WilMau+92,NehGry+08}.  As well, the
identification of absorbing gas in CO emission would offer the
prospect of relatively high spatial resolution imaging of the
absorption line host body.  This being the case, 
it seemed reasonable to survey CO J=1-0 rotational emission 
toward a substantial sample of commonly-observed optical/{\it uv} 
absorption sightlines.  A small but nominally similar 
survey was performed at a much earlier epoch by \cite{KnaJur76}, 
with somewhat mixed results owing to confusion between telluric and
interstellar emission ( Appendix A).   The results 
are somewhat mixed here as well owing to confusion between
foreground and background material (see \cite{NehGry+08}), 
combined with the propensity for early-type absorption line target stars 
to be associated with but in front of large amounts of neutral 
material as discussed in Sect. 4 and 5.

Section 2 here describes the data which were taken to implement the survey
and the methods of sightline selection, data reduction and presentation.  
Section 3 describes the observational results. Section 4 reports the survey
statistics and discusses some evident biases.    Section 5 discusses the
apparent difficulties in separating foreground and background gas in 
emission and Sect. 6 is a summary.  Appendices A-C discuss various 
observational aspects of the implementation of the
present survey, including telluric CO emission, spectral baseline
removal and deconvolution of frequency-switching (respectively) .


\section{Sample selection and overlap with prior work}

\begin{table}
\caption[]{Sightlines lacking interstellar CO J=1-0 emission$^1$}
{
\begin{tabular}{lcccccc}
\hline
Target& l & b & v$_{\rm tel}$ & rms &  N(CO) & N(\HH) \\
HD &$\deg$ &$\deg$ & \kms\ & K & $\pcc$ & $\pcc$  \\
\hline
13994&134.39 & -3.42 &-6.17 &0.098   && \\
21856&196.46 &-50.96 &-29.36 &0.124   && 20.04\\
22951&158.92 &-16.70 &-14.08 &0.087  &14.22 &20.46 \\
23408&166.17 &-23.51 &-18.11 &0.098   && 19.75\\
23480&166.57 &-23.75 &-17.98 &0.169   &&20.12 \\
23630&166.67 &-23.46 &-18.32 &0.088& $<$12.34 &19.54   \\
23850&167.01 &-23.23 &-18.56 &0.162   && \\
23840&159.46 &-15.03 &-33.41 &0.055    &&\\
24912&160.37 &-13.11 &-12.78 &0.066 &13.49 &20.53  \\
30614&144.07 & 14.04 &-14.38 &0.073  &14.49 &20.34 \\
34078&172.08 & -2.26 &-6.72 &0.126    &&\\
35149&199.16 &-17.86 &-14.71 &0.107& $<$13.00 &18.30   \\
36486&203.86 &-17.74 &-14.17 &0.116 & $<$12.04 &14.74 \\
37128&205.21 &-17.24 &-14.13 &0.085 & $<$12.30 &16.28  \\
42087&187.75 &  1.77 &-30.60 &0.098   && \\
57060&237.82 & -5.37 &-5.79 &0.157  & $<$12.67 &15.78 \\
57061&238.18 & -5.54 &-5.89 &0.158  & $<$12.61 &15.48 \\
58510&235.52 & -2.47 &-23.78 &0.138   && \\
121968&333.97 & 55.84 &29.95 &0.145  & $<$12.30 &18.70\\
143275&350.10 & 22.49 &18.28 &0.099  &12.49 &19.41 \\
144217&353.19 & 23.60 &35.21 &0.116 &13.63 &19.83 \\
144470&352.75 & 22.77 &34.93 &0.120  &12.95 &20.05\\
145502&354.61 & 22.70 &34.88 &0.120  &12.76 &19.89 \\
164353& 29.73 & 12.63 &10.47 &0.086  &13.04 &20.26 \\
166937& 10.00 & -1.60 &24.39 &0.131   && \\
167971& 18.25 &  1.68 &5.93 &0.129   && \\
170074& 75.06 & 23.71 &14.80 &0.090   && \\
177989& 17.81 &-11.88 &-1.67 &0.123 &14.64 &20.23 \\
181615& 21.84 &-13.77 &-2.31 &0.153   && \\
184915& 31.77 &-13.29 &-2.70 &0.101   && 20.31\\
192639& 74.90 &  1.48 &2.07 &0.131 &14.13 &20.69   \\
193237& 75.83 &  1.32 &2.23 &0.102    &&\\
197592& 30.70 &-32.15 &-13.21 &0.121   && \\
198478& 85.75 &  1.49 &1.09 &0.102   && \\
198781& 99.94 & 12.61 &6.23 &0.066&15.22 &20.56  \\
199579& 85.70 & -0.30 &0.23 &0.112 && 20.36 \\
201345& 78.44 & -9.54 &-4.69 &0.064 & $<$12.40 &19.43 \\
203064& 87.61 & -3.84 &-1.71 &0.092   && 20.29\\
206773& 99.80 &  3.62 &0.55 &0.101 &14.20 &20.47 \\
209339&104.58 &  5.87 &1.96 &0.105  &13.95 &20.25  \\
209975&104.87 &  5.39 &1.80 &0.080   && 20.08\\
217035&110.25 &  2.86 &-0.26 &0.090 &14.57 &20.95 \\
217675&102.21 &-16.10 &-10.33 &0.135 &12.75 &19.67  \\
218915&108.06 & -6.89 &-5.68 &0.134 &13.64 &20.15 \\
224572&115.55 & -6.36 &-5.99 &0.113   && \\
\hline
\end{tabular}}
\\
$^1$For explanation of table entries see Sect. 3 \\
All column densities are logarithmic. \\
\end{table}

The goal of this work was to observe as many targets as possible given 
prior absorption line surveys of important species \HH, CO, OH, CH, 
CH\p\ and \H3\p.  Owing to terrestrial geography, only targets above 
-25\degr\ declination were included;
given that absorption lines of \HH\ and CO are observed from space and
that that the center of the Galaxy is in the southern sky, this criterion 
eliminated many sightlines. The final target list numbered 110 and two 
sightlines near Orion B were added to the observational program
(but excluded from calculation of the statistics) as controlled examples 
of heated gas along sightlines to well-studied HII regions 
\citep{PetGoi+07}. 

First priority was given to  sightlines observed in $uv$ CO absorption by
\cite{SonWel+07}, \cite{BurFra+07} and \cite{SheRog+07}.  These references 
also tabulate or provide values or new measurements of N(\HH).  
Sightlines having measured 
N(\HH) not included in the CO surveys were taken from the work of 
\cite{SavDra+77} and \cite{RacSno+02}.  CH column densities for 
sources surveyed in CO absorption are included in the work of 
\cite{SonWel+07} and some further sightlines studied in CH 
were taken from the work of \cite{CraLam+95} \footnote{Several of the 
lines of sight in the work of \cite{CraLam+95} are denoted by 
their HR numbers, specifically HR1156 = HD23480, HR1178 = HD23850,
HR2135 = HD41117 and HR6812 = HD166937}.  Finally, a few sightlines
were included which have recently between observed in CH\p\
by \cite{StaCas+08} (CH\p\ was also surveyed by \cite{CraLam+95})
and in H$_3$\p\ absorption by \cite{McCHin+02}.

\section{Observations and data reduction}

The new data discussed here, profiles of J=1-0 \cotw\ and \coth\ emission,
were acquired at the ARO 12m telescope during 2007 December and to a lesser 
extent in 2008 January and February, to make up for time lost to weather 
and objects too close to the Sun earlier.  The observations are somewhat
time-specific because of the presence of telluric emission in the \cotw\
spectra.   At the frequency of the \cotw\ J=1-0 line, 115.3 GHz,
the half-power beamwidth of the 12m antenna is 65\arcsec.  Line brightnesses
from the 12m telescope are on a \Tstar\ scale and should be scaled upward
by a factor 0.85/0.72 = 1.2 to derive corresponding main-beam brightness. 

In an all-sky survey of this sort it is not practicable to search for nearby 
off-source positions which are free of emission at low levels.  Therefore,
the raw data were taken in a frequency-switching mode \citep{Lis97FS} with the 
6144-channel mm-wave autocorrelator (MAC) at 24.4 kHz resolution (0.0635 \kms\
at the \cotw\ rest frequency).  The throw of the frequency-switch was $\pm1$ MHz
for most spectra and  $\pm2$ MHz for a few others which at  $\pm1$ MHz were hard to 
reconstruct faithfully. Typical integration times were 12-18 minutes in each of two 
polarizations resulting in a typical rms of 0.1 K after the spectra were 
co-added in the two polarizations and hanning-smoothed to a final resolution 
of 48.8 kHz (0.127 \kms\ for \cotw).

The baselines in
frequency-switched data at some mm-wave telescopes are bad enough to prevent
reliable reconstruction (unfolding) of broad-band spectra, but those at the 12m 
were sums of two or three pure sine-waves, typically dominated by 
periods of approximately 156 and 17.4 MHz (400 and 35 \kms, respectively for 
\cotw), making baseline subtraction simple.  This is illustrated in Appendix B
and Fig. A.2.  

The data were unfolded using the EKHL algorithm \citep{Lis97FS}  illustrated in
Fig. A.3 and discussed in Appendix C.  This technique, 
analogous to the dual-beam switching
procedure for spatial mapping, allows recovery of spectra with frequency-switch
intervals smaller than the extent of the emission profile, as long as the spectral 
baselines are flat.  In turn, because the switching interval may be made smaller,
the baselines may also improve.  There is a penalty to be paid in terms of
time because the rms noise may not decrease much after unfolding, but this
is negligible compared to the amount of time which would be spent hunting for 
a suitable off-source reference position (which would be followed by observing 
in a position-switching mode, with less than half the observing time spent 
on-source anyway).  The penalty is also negligible compared to the process
detailed in Fig. 8 of \cite{McCHin+02} where spectra taken with 3 frequency
throws toward HD183143 did not yield a spectrum showing all
the emission present along the line of sight (see Fig. 2 here).

\begin{table*}
\caption[]{Sightlines with detected interstellar CO J=1-0 emission$^1$}
{
\begin{tabular}{lcccccccccccc}
\hline
Target & l & b & \EBVl & \EBV &v$_{\rm tel}$ & rms & \Tstar & W$_{\rm CO}$ & 
  $< {\rm V_{\rm lsr}} >$  & V$_{\rm lsr}$ - V$_{\rm hel}$ 
 & N(CO) & N(\HH)  \\
 &$\deg$ &$\deg$ & mag & mag & \kms\ & K & K & K \kms\ & \kms\ & \kms\ & $\pcc$ & $\pcc$  \\
\hline
HD2905 & 120.84 &   0.14 &  1.90 & 0.01 &  -2.72 & 0.179 &  4.12 &  8.40 & -52.42 &   7.86 & & 20.27  \\
HD21483 & 158.87 & -21.30 &  1.04 & 0.56 & -16.46 & 0.083 &  7.59 & 22.76 &   6.06 &  -6.41&& \\
HD23180 & 160.36 & -17.74 &  4.54 & 0.30 & -14.71 & 0.108 & 11.46 & 19.62 &   8.51 &  -6.49 & 14.83 & 20.60  \\
HD281159 & 160.49 & -17.80 & 10.02 & 0.85 & -14.74 & 0.059 & 27.97 & 41.98 &   8.35 &  -6.54&& \\
HD24398 & 162.29 & -16.69 &  0.32 & 0.32 & -14.61 & 0.086 &  0.97 &  2.13 &   8.05 &  -6.95 & 14.86 & 20.67  \\
HD24534 & 163.08 & -17.14 &  0.43 & 0.45 & -14.92 & 0.073 &  2.55 &  5.31 &   7.60 &  -7.23 & 16.20 & 20.92  \\
HD24760 & 157.35 & -10.09 &  0.95 &  & -10.71 & 0.074 &  4.57 & 22.68 &  -5.45 &  -4.75 & 11.93 & 19.52  \\
HD27778 & 172.76 & -17.39 &  0.60 & 0.37 & -15.24 & 0.080 &  3.38 &  7.17 &   5.74 &  -9.98 & 16.08 & 20.79  \\
HD29647 & 174.05 & -13.35 &  2.31 & 1.00 & -13.09 & 0.090 &  5.84 & 16.35 &   6.14 &  -9.98&& \\
HD283809 & 174.16 & -13.35 &  2.33 &  & -13.11 & 0.114 &  5.08 & 16.74 &   5.95 & -10.01&& \\
HD283879 & 175.72 & -12.61 &  1.50 &  & -12.71 & 0.095 &  6.79 & 13.23 &   5.92 & -10.36&& \\
HD36371 & 175.77 &  -0.61 &  1.54 & 0.43 &  -6.29 & 0.137 &  1.26 &  3.60 & -18.90 &  -9.04&& \\
HD36861 & 195.05 & -11.99 &  0.45 & 0.12 & -11.62 & 0.119 &  1.22 &  1.13 &   5.12 & -14.95 & 12.59 & 19.12 \\
HD37043 & 209.52 & -19.58 &  1.51 & 0.07 & -15.17 & 0.106 & 20.15 & 58.57 &   8.01 & -17.83 & 12.43 & 14.69  \\
HD37742 & 206.45 & -16.59 &  4.20 & 0.08 & -13.88 & 0.096 &  5.42 & 18.58 &  10.06 & -17.28 & $<$12.72 & 15.86  \\
HorseHead & 206.96 & -16.79 &  0.98 &  &   0.00 & 0.069 & 15.92 & 31.80 &   9.79 & -17.37&& \\
NGC2024\#1 & 206.53 & -16.38 & 58.85 &  & -37.16 & 0.064 & 31.46 &156.18 &   9.95 & -17.28&& \\
HD37903 & 206.85 & -16.54 &  2.05 & 0.35 & -37.35 & 0.069 & 41.20 &134.38 &   9.46 & -17.34 & 14.00 & 20.92  \\
HD41117 & 189.69 &  -0.86 &  1.67 & 0.45 & -32.27 & 0.050 &  1.75 &  3.54 &   4.51 & -12.62&& \\
HD43384 & 187.99 &   3.53 &  0.81 & 0.58 & -29.67 & 0.114 &  1.31 &  1.19 &   2.41 & -11.63&&\\
HD46202 & 206.31 &  -2.00 &  0.82 & 0.49 & -31.43 & 0.102 &  1.38 &  0.32 &   9.53 & -16.03&& \\
HD46711 & 208.59 &  -2.40 &  2.17 & 1.04 & -31.02 & 0.103 &  1.75 &  2.74 &  23.23 & -16.43&& \\
HD47129 & 205.87 &  -0.31 &  0.72 & 0.36 & -30.65 & 0.099 &  1.35 &  1.17 &  20.08 & -15.75 & & 20.54  \\
HD47839 & 202.94 &   2.20 &  6.31 & 0.07 &  -3.21 & 0.090 & 12.45 & 47.99 &   6.87 & -14.92 & $<$12.81 & 15.54  \\
HD48099 & 206.21 &   0.80 &  1.13 & 0.27 & -30.18 & 0.095 &  1.49 &  1.21 &  15.98 & -15.66 & & 20.29  \\
HD52382 & 222.17 &  -2.15 &  1.99 &  & -28.05 & 0.123 &  2.45 &  4.86 &  39.30 & -17.94&& \\
HD53367 & 223.71 &  -1.90 &  6.89 & 0.74 & -27.43 & 0.109 & 13.32 & 47.45 &  17.22 & -18.02&& \\
HD53974 & 224.71 &  -1.79 &  1.99 & 0.22 & -27.37 & 0.067 &  5.29 & 12.55 &  15.48 & -18.07&& \\
HD54662 & 224.17 &  -0.78 &  1.61 & 0.35 & -27.21 & 0.062 &  4.21 & 16.63 &  14.19 & -17.91 & & 20.00  \\
HD147165 & 351.31 &  17.00 &  1.11 & 0.38 &  32.57 & 0.113 &  2.16 &  0.46 &   3.12 &   9.55 & 12.48 & 19.79   \\
HD147888 & 353.65 &  17.71 &  1.26 & 0.47 &  33.10 & 0.112 &  3.14 &  3.49 &   2.71 &  10.25 & 15.30 & 20.48   \\
HD147889 & 352.86 &  17.04 & 15.13 & 1.07 &  32.76 & 0.122 & 16.74 & 30.86 &   3.13 &   9.98&& \\
HD147933 & 353.69 &  17.69 &  1.23 & 0.45 &  32.85 & 0.124 &  2.93 &  3.39 &   2.63 &  10.25 & 15.28 & 20.57   \\
HD148184 & 357.93 &  20.68 &  0.49 & 0.55 &  34.41 & 0.105 &  1.80 &  1.10 &   1.14 &  11.58 & 14.58 & 20.63   \\
HD148605 & 353.10 &  15.80 &  2.03 & 0.10 &  32.26 & 0.123 &  2.04 &  0.99 &   3.67 &   9.94 & & 18.74  \\
\hline
\end{tabular}}
\\
$^1$ For explanation of table entries see Sect. 3 . All column densities are logarithmic. \\
\end{table*}

\begin{table*}
\caption[]{Sightlines with detected interstellar CO J=1-0 emission$^1$}
{
\begin{tabular}{lcccccccccccc}
\hline
Target & l & b & \EBVl & \EBV & v$_{\rm tel}$ & rms & \Tstar & W$_{\rm CO}$ & 
  $< {\rm V_{\rm lsr}} >$  & V$_{\rm lsr}$ - V$_{\rm hel}$ 
 & N(CO) & N(\HH)  \\
 &$\deg$ &$\deg$ & mag & & mag \kms\ & K & K & K \kms\ & \kms\ & \kms\ & $\pcc$ & $\pcc$ \\
\hline
HD149757 &   6.28 &  23.59 &  0.56 & 0.32 &  35.83 & 0.102 &  1.60 &  1.66 &  -0.07 &  13.72 & 15.40 & 20.64  \\
HD166734 &  18.92 &   3.63 &  1.52 & 1.39 &   7.74 & 0.094 &  2.43 &  3.90 &   6.15 &  14.92&& \\
HD168076 &  16.94 &   0.84 & 25.94 & 0.78 &   6.29 & 0.094 &  5.99 & 38.82 &   0.88 &  14.19&& \\
HD168607 &  14.97 &  -0.94 &  8.07 & 1.61 &   5.16 & 0.165 &  4.77 & 74.31 &  38.38 &  13.55&& \\
HD169454 &  17.54 &  -0.67 & 13.94 & 1.12 &   4.35 & 0.222 &  7.72 & 12.12 &   5.55 &  14.12&&  \\
HD172028 &  31.05 &   2.83 &  2.97 & 0.78 &   6.45 & 0.118 &  7.65 & 23.03 &   7.58 &  16.83&& \\
HD179406 &  28.23 &  -8.31 &  0.49 & 0.33 &  -1.34 & 0.083 &  1.26 &  3.09 &   2.46 &  14.82 & 15.64 & 20.73   \\
HD183143 &  53.24 &   0.63 &  3.75 & 1.27 &   2.80 & 0.129 &  3.03 & 15.90 & -22.61 &  18.28&& \\
HD192035 &  83.33 &   7.76 &  0.46 & 0.28 &   4.86 & 0.071 &  1.16 &  2.13 &   5.58 &  17.10 & 15.14 & 20.68   \\
HD195592 &  82.36 &   2.96 &  2.75 & 1.18 &   2.34 & 0.062 &  1.14 &  3.85 &   2.28 &  16.74&& \\
CygOB2\#8 &  79.89 &   0.96 &  5.83 &  &   1.11 & 0.057 &  4.95 & 50.36 & -66.24 &  16.82&& \\
CygOB2\#5 &  80.12 &   0.91 &  3.46 & 1.99 &   1.33 & 0.072 &  2.01 & 23.26 &  -3.90 &  16.79&& \\
CygOB2\#12 &  80.10 &   0.83 &  4.18 & 3.31 &   0.97 & 0.063 &  4.13 & 27.44 &   1.93 &  16.78&& \\
VICYG\#8A &  80.22 &   0.79 &  6.46 &  &   1.04 & 0.062 &  3.63 & 27.49 & -10.74 &  16.76&& \\
HD200120 &  88.03 &   0.97 &  2.67 & 0.18 &   0.62 & 0.061 &  1.76 &  9.69 & -38.87 &  15.63 & 12.18 & 19.30  \\
HD203374 & 100.51 &   8.62 &  1.09 & 0.60 &   3.89 & 0.076 &  1.18 &  3.19 &   2.20 &  14.04 & 15.41 & 20.70  \\
HD203938 &  90.56 &  -2.23 &  1.55 & 0.74 &  -1.36 & 0.073 &  2.90 &  4.97 &   5.15 &  14.76&& \\
HD204827 &  99.17 &   5.55 &  1.05 & 1.11 &   1.39 & 0.088 &  2.81 & 11.21 & -66.86 &  14.02&& \\
HD206165 & 102.27 &   7.25 &  0.70 & 0.47 &   2.96 & 0.090 &  1.27 &  1.86 &  -0.36 &  13.51 & 15.18 & 20.78   \\
HD206267 &  99.29 &   3.74 &  0.98 & 0.52 &   1.19 & 0.084 &  1.54 &  2.91 &  -2.24 &  13.79 & 16.04 & 20.86   \\
HD207198 & 103.14 &   6.99 &  0.66 & 0.62 &   2.67 & 0.153 &  1.61 &  2.63 & -26.88 &  13.29 & 15.50 & 20.83   \\
HD207308 & 103.11 &   6.82 &  0.59 & 0.50 &   2.75 & 0.123 &  1.88 &  4.31 &  -2.18 &  13.28 & 15.93 & 20.86   \\
HD207538 & 101.60 &   4.67 &  1.19 & 0.64 &   1.27 & 0.082 &  1.33 &  2.19 &  -1.89 &  13.39 & 15.40 & 20.91   \\
HD208440 & 104.03 &   6.44 &  0.66 & 0.34 &   2.28 & 0.133 &  1.22 &  1.10 &  -1.06 &  13.02 & 14.20 & 20.34   \\
HD210121 &  56.88 & -44.46 &  0.22 & 0.40 & -21.62 & 0.140 &  2.00 &  2.91 &  -6.32 &   7.76 & 15.83 & 20.75   \\
HD210839 & 103.83 &   2.61 &  1.64 & 0.56 &  -0.12 & 0.107 &  1.14 &  4.16 & -77.96 &  12.64 & 15.44 & 20.84   \\
HD216532 & 109.65 &   2.68 &  3.73 & 0.86 &  -0.11 & 0.085 &  2.35 &  9.00 & -98.72 &  11.22 & 15.15 & 21.10   \\
HD216898 & 109.93 &   2.39 &  2.86 & 0.85 &  -0.66 & 0.084 &  0.97 &  4.59 &  -9.89 &  11.11 & 15.04 & 21.05   \\
HD217312 & 110.56 &   2.95 &  1.04 & 0.66 &   0.07 & 0.154 &  1.60 &  3.09 & -78.91 &  11.02 & 14.38 & 20.80   \\
HD218376 & 109.95 &  -0.78 &  1.51 & 0.25 &  -2.14 & 0.138 &  1.47 &  7.58 & -45.05 &  10.70 & 14.80 & 20.15   \\
BD+63 1964 & 112.89 &   3.10 &  1.70 & 1.00 &  -0.05 & 0.115 &  1.38 &  2.73 & -10.86 &  10.43&& \\
HD229059 &  75.70 &   0.40 & 13.52 & 1.71 &   1.05 & 0.055 & 10.00 & 74.64 & -52.72 &  17.24&& \\
\hline
\end{tabular}}
\\
$^1$For explanation of table entries see Sect. 3 . All column densities are logarithmic \\
\end{table*}

The velocity scale of all the spectra is with respect to the LSR. The data 
output from the 12m are on a \Tstar\ scale, that is, corrected
for a beam efficiency of 0.85 corresponding to the forward response
falling within the area of the Moon.  To put the interstellar data on 
a main-beam scale, the spectra should be scaled up by a factor 
$\approx$ 1.2 (0.85/0.7) corresponding to the full main beam efficiency. 

The survey results are summarized in Tables 1-4.  Table 1 describes
sightlines lacking detected interstellar emission; it provides
the galactic coordinates, the (date-specific) mean lsr velocity at 
which  telluric emission appeared in the spectra 
(labelled  v$_{\rm tel}$), the channel to channel rms of the 
reduced, coadded and smoothed ARO spectrum and the 
 log of the absorbing 
CO and \HH\ column densities known from other work.  Tables 2 
and 3 summarize the lines of sight with detected interstellar CO.   
For each source they provide galactic coordinates,
the reddening out to infinity from the work of \cite{SchFin+98}, 
 the foreground reddening copied from earlier references,
 v$_{\rm tel}$, followed by the peak and integrated \Tstar, 
the mean velocity of the emission (which in a few cases contains
an indistinguishable  telluric contribution), the correction 
V$_{\rm lsr}$ - V$_{\rm hel}$ which may be subtracted
from the lsr velocity scale to convert to heliocentric velocity, 
followed by the  log of the  absorption line column densities of CO and \HH. 
Table 4 provides relevant information for the 14 lines of sight where
\coth\ was also observed.  Integration times for these lines of sight
were somewhat longer, typically 24 minutes.

Thumbnail plots of the spectra with detectable interstellar CO emission
are shown in Figures 1-3; where \coth\ was observed, those spectra are 
shown overlaid.  When $|{\rm v}_{\rm tel}| > 10-15$ 
\kms\ and the galactic latitude is a few degrees or more from the galactic
plane, overlap with interstellar emission is much less likely.  
Conversely, there are examples where the telluric
line inextricably appears within the velocity range of the interstellar
emission and the spectra can only be understood in complete detail
with reference to the tabulated telluric velocities v$_{\rm tel}$.   
In general the telluric and interstellar emission can 
be separated by observing at several-month intervals.

\subsection{Data in machine-readable form}

Individual spectra and zip files of all spectra having detected
interstellar CO emission are available in CLASS FITS format on 
NRAO's anonymous ftp server as http://tinyurl.com/45ps73 and 
http://tinyurl.com/5pgju9 for \cotw\ and \coth\ spectra, respectively.

\begin{figure*}
\psfig{figure=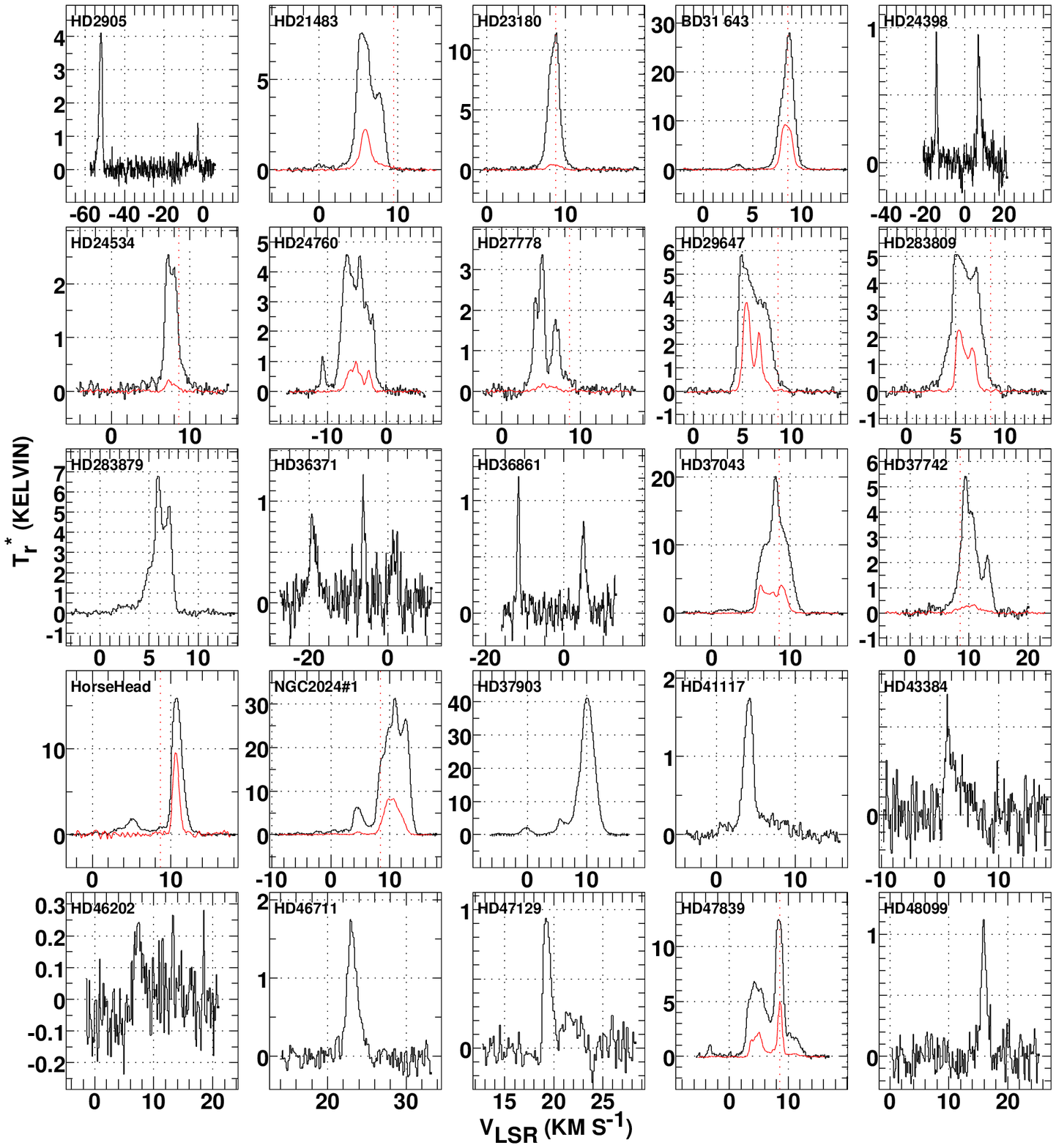,height=19cm}
\caption[]{Observed CO J=1-0 emission profiles.  
Velocity resolution is generally 0.13 \kms\ but a few more extensive 
profiles have been smoothed to 0.25 \kms}
\end{figure*}

\begin{figure*}
\psfig{figure=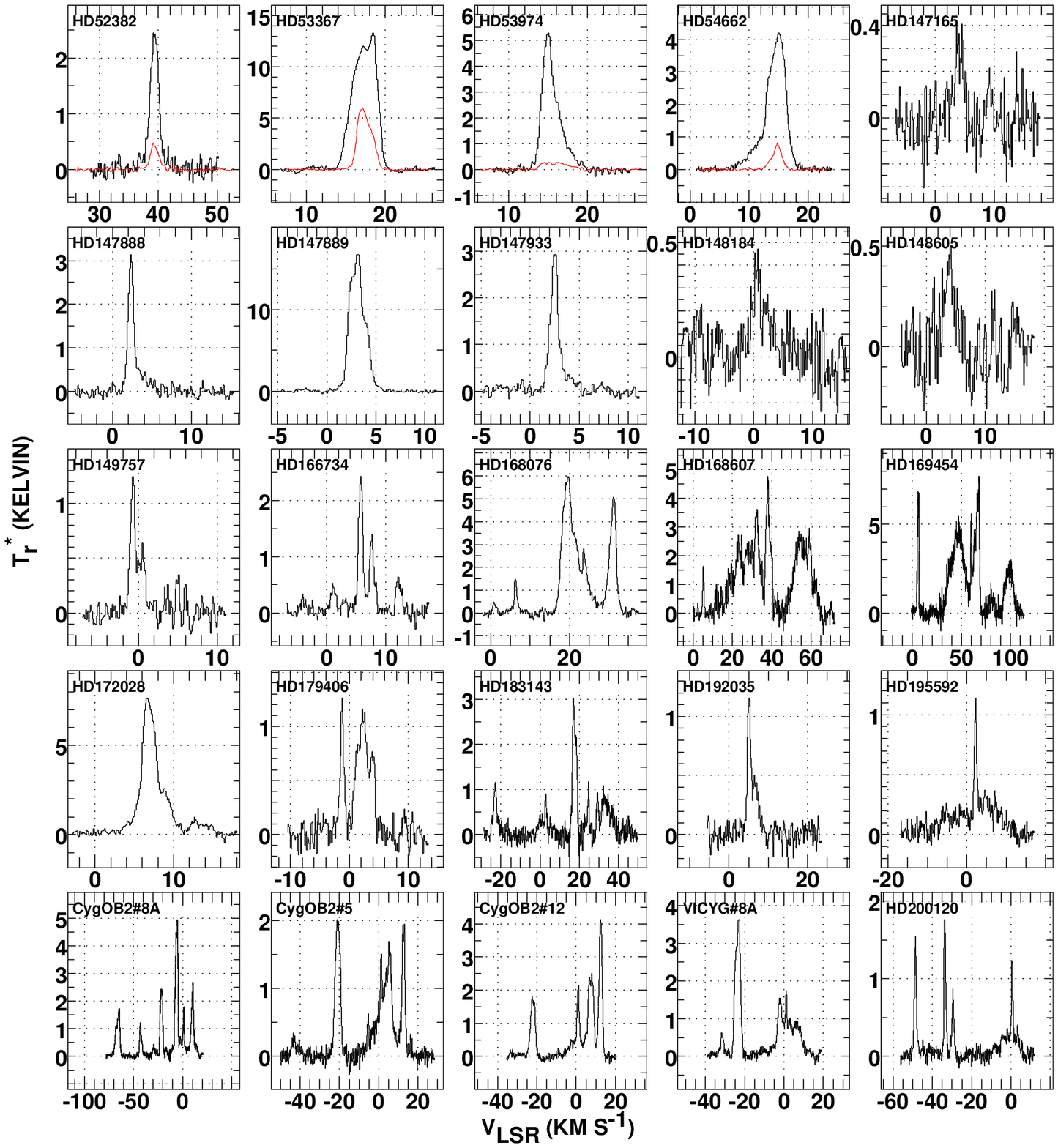,height=19cm}
\caption[]{Observed CO J=1-0 emission profiles, as in Fig. 1}
\end{figure*}

\begin{figure*}
\psfig{figure=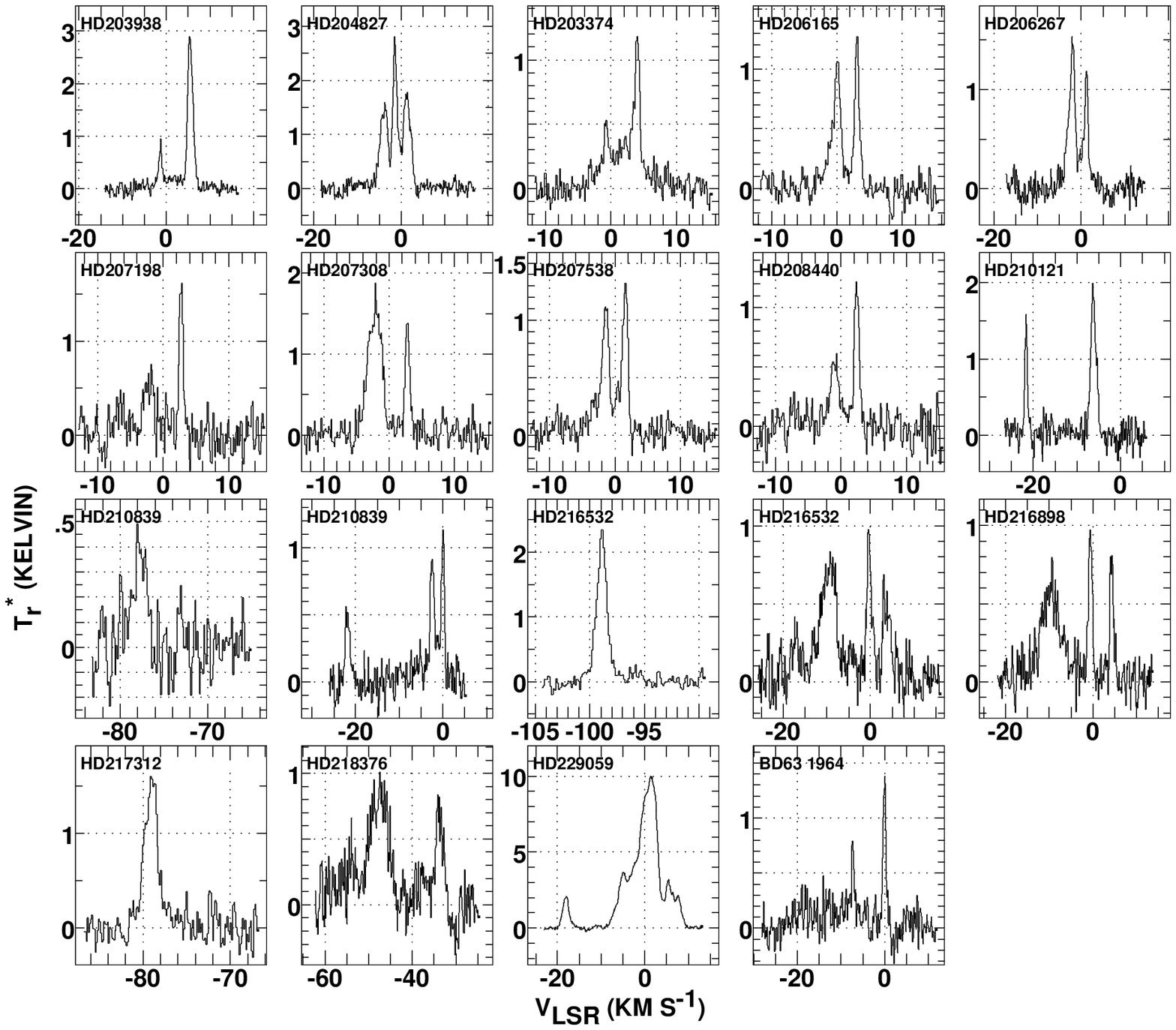,height=15cm}
\caption[]{Observed CO J=1-0 emission profiles, as in Fig. 1}
\end{figure*}

\subsection{Detectability of interstellar emission and presentation of upper
limits}

The spectra acquired here easily detect telluric emission which appears
with \WCO\ $\approx 1$ K \kms/sin(el) and FWHM = 0.8 \kms\ (see Appendix A); 
the nominal signal/noise
of such a detection is nearly 20:1 for a channel-to-channel rms of 0.1 K.  
For a 4 \kms\ interval, typical of what is needed to bracket a
single hypothetical interstellar component, the 3$\sigma$ rms noise in 
\WCO\ with single channel rms noise 0.1 K is 0.22 K \kms.  As discussed
in Sect. 5 (see Fig. 5 at bottom), interstellar emission was detected in 
every case, save one, where N(CO) exceeded $ 4\times 10^{14} \pcc$ 
and emission was expected at a level \WCO\ $>$ 0.4 K \kms.  However, both 
individual components and lines of sight with detected interstellar lines 
at levels \WCO\ $<$ 1 K \kms\ are rare.  Because no threshold for CO emission 
is expected empirically  (\WCO\ $\propto$ N(CO), see Fig. 6 of
\cite{Lis07CO}),  the actual {\it a posteriori} sensitivity of 
the survey may be somewhat poorer than suggested by the detectability
of the narrow telluric emission or the 0.1 K single channel rms noise.
Where upper limits appear in the Figures (4 and 5), they are shown symbolically
at \WCO\ $<$ 0.8 K \kms.  The non-detections in Table 1 should be reliable
at a level \WCO\ $<$ 0.5 - 0.8 K \kms.

\section{Survey results}

\subsection{Statistics and biases}

Given that the purpose of the survey (which was achieved) was to 
provide line profiles toward a pre-defined but rather {\it ad hoc} 
group of target sightlines, the statistical properties of the CO 
emission found in the survey might 
not seem to be of much intrinsic value.  However, they serve to show 
that the targets are associated with the gas detected (making the gas 
atypical of the diffuse ISM as a whole) and that CO emission 
surveys which seek the foreground gas contribution toward bright stars
are heavily contaminated (if not totally corrupted) by 
material lying behind the absorption targets.  

This contamination occurs for several reasons.  First is the innate 
inability of the emission spectra to discriminate between foreground 
and background placement of material near the target in the absence 
of other information (for instance, but not necessarily conclusively, 
the velocity profile of the absorbing gas).  Second is the fact that 
when substantial amounts of gas {\it are} present along such lines 
of sight, they will preferentially occur behind the stars, which
were {\it a priori} selected to observe the diffuse cloud regime
\AV\ $\la 1$ mag.   Finally, this bias is greatly enhanced by 
the propensity for the target stars to occur in regions of high 
gas density, presumably because they are generally of early spectral 
type and were deemed to be ``interesting'' for spectroscopic studies.

The association between the targets and the CO emitting gas can be 
demonstrated in several ways.  It is obvious from the spectra that 
several of the sources (HD23180, HD281159, HD37043, HD37903, HD147889)
are seen against very strongly-emitting molecular gas characteristic
of natal clouds near giant HII regions, while others among the few
observed in \coth\ show the optically thick CO emission typical of
dark clouds (HD21483, HD24760, HD29647, HD283809).  H II regions 
and dark clouds are rarely found by accident away from the galactic 
equator.   In any case, the strong CO lines indicate very high column 
densities of material behind which the stars would not be suitable 
targets for optical/$uv$ absorption line studies of diffuse clouds.

The existence of an overendowment of gas along the chosen lines of sight 
is apparent in at least three ways:  the amount of reddening along the 
ensemble of lines of sight is much higher than expected; CO is detected too 
frequently and the CO emission is on average too strong.  Each
of these is discussed separately in the following subsections.

\subsection{Reddening and the total gas column}

\begin{table}
\caption[]{Sightlines observed in \coth$^1$}
{
\begin{tabular}{lccccc}
\hline
Target & rms & \Tstar & W$_{13}$ &  $< {\rm V_{\rm lsr}} >$ & W$_{12}$/W$_{13}$ \\
HD &K & K & K \kms\ & \kms\ & \\
\hline
21483&0.038&2.23&3.47&5.93& 6.56 \\
23180&0.025&0.44&0.84&8.37& 23.46 \\
281159&0.033&9.20&13.38&8.61& 3.14 \\
24534&0.017&0.22&0.29&7.77& 18.49 \\
24760&0.023&1.01&3.02&-4.97& 7.52 \\
27778&0.021&0.19&0.43&5.96& 16.86 \\
29647&0.031&3.79&5.84&5.88& 2.80 \\
283809&0.025&2.27&4.05&5.96& 4.13 \\
37043&0.038&4.14&13.92&7.87& 4.21 \\
37742&0.030&0.33&0.95&10.10& 19.52 \\
47839&0.027&4.98&10.14&7.00& 4.73 \\
52382&0.024&0.48&0.63&39.32& 7.71 \\
53367&0.047&5.94&13.98&17.41&3.39 \\
53974&0.022&0.29&0.91&16.15& 13.82 \\
\hline 
\end{tabular}}
\\
$^1$For explanation of table entries see Sect. 3 \\
\end{table}


 Table 5 gives some statistics of the reddening and integrated
CO emission.  For a series of minimum separations from the galactic 
equator $|{\rm b}|_< = 3$\degr - 20\degr, the columns of Table 5 show
the subsumed number of lines of sight; the number among them
with \EBVl\ $> 1$ mag;  the mean reddening per kpc of equivalent path 
in the galactic plane $<\EBVl>/<l>$, where $l$ =  0.1 kpc/sin($|$b$|$) 
corresponding to a half-thickness of the dust layer of 100 pc; the number
of sightlines with detected CO emission and mean integrated CO emission per
unit length (calculated as for the reddening per unit length); the ratio
of the reddening and CO emission per unit length, converted to an equivalent 
column density of \HH; and finally, four columns giving the mean foreground 
and limiting reddening, tabulated separately for sightlines with and without
detectable CO emission.  To eliminate the influence of a few outliers (see Tables
2-3), the mean reddening per unit length was calculated using sightlines
with \EBVl $<$ 5 mag and the mean CO emission per unit length was calculated
for \WCO $<$ 50 K \kms.  The table entries for either quantity would have
been 50-75\% higher if these limits were not observed.

For a local mean reddening at z = 0 of 
0.61 mag kpc$^{-1}$ \citep{Spi78}, lines of sight with \EBVl\ $>$ 1 mag 
should be confined within about 4\degr\ of the galactic plane, 
which is manifestly not the case here.  Moreover, the mean  limiting
reddening per unit path in Table 5 ranges from 2 to 5 times the local 
average depending on the lower latitude limit.  This
can be contrasted with the opposing tendency for absorption line
surveys to have smaller than average reddening per unit length when
only the foreground material is counted \citep{SavDra+77}.

\subsection{Frequency of occurence and strength of CO emission}


In the present survey, 31 of the 55 lines of sight at 3\degr\ $< b < 20$\degr\ 
had detectable emission (56\%) compared to 10 of 48 lines of sight 
(22\%) in the same latitude range in a truly blind survey for galactic 
CO emission toward extragalactic mm-wave radiocontinuum sources 
\citep{LisWil93,Lis94}.  Furthermore, the emission lines found in the 
blind survey were much weaker, illustrating that the present work 
also includes a superabundance of integrated CO emission (not just 
detections) at high latitude.   The mean integrated brightnesses per 
kpc of path in this survey at $|$b$| > 3$\degr\ are
\ACO\ = 14-39 K \kms\ kpc$^{-1}$, generally increasing with the 
lower bound of the latitude range considered.  This may be compared 
with a value \ACO\ $\approx$  5 K  \kms\ kpc$^{-1}$ at the Solar Circle 
in surveys of the galactic plane \citep{BurGor78}.

\subsection{Foreground {\it vs.} background reddening}

 The final four columns of Table 5 give the mean foreground and limiting 
reddening for sightlines with and without detected CO emission.  The 
sample mean foreground reddening for the 71 lines of sight is 0.35 mag
corresponding to \AV\ $\approx 3.1 $\EBV\ = 1.1 mag, observing the diffuse 
cloud regime \AV $\la$ 1 mag. Those sightlines showing CO emission have a 
somewhat larger overall mean foreground reddening, 0.48 {\it vs.} 0.21 mag 
and a much larger mean limiting reddening, 1.87 {\it vs.} 0.48 mag.  
This is a quantitative demonstration of the bias noted earlier whereby 
associated material preferentially occurs behind early-type target stars 
whose foreground extinction has been {\it a priori} selected to fall 
in the diffuse cloud regime.  Moreover, CO emission is preferentially
detected in the more extreme cases.

\subsection{Relationships between \WCO\ and \EBVl}

The endowments of limiting reddening and CO emission are generally 
proportional and their ratio, converted to an equivalent column density 
of \HH\ in Table 5 following \cite{SavDra+77},  is relatively constant.
It is also comparable to typical values of the CO-\HH\ conversion factor 
used to convert CO intensity to \HH\ column density, 
$2\times10^{20}$ \HH (K \kms)$^{-1}$, suggesting that the material
represented by \EBVl\ is largely in molecular form for the larger
values of \EBVl\ which dominate the ensemble mean.
  

Line of sight effects are expressed differently in Fig. 4.  At  top,
Fig. 4 plots the distribution of \WCO\ with csc($|$b$|$); as observed
in the blind surveys for CO emission toward extragalactic contiuum
sources \citep{LisWil93,Lis94}, there is no plane-parallel stratification 
of the CO-emitting medium.  Unlike in the blind survey, the reason now
is not the galactic structure of the local bubble but rather the association 
of CO-emitting gas with the target stars.

At  bottom in Fig. 4, \WCO\ is plotted against the limiting 
line of sight reddening out to infinity \EBVl\ from \cite{SchFin+98}.  
As suggested by the near constancy of \WCO/\EBVl\ in Table 5 
there is a substantial subset of lines of sight which show a 
proportionality between \WCO\ and \EBVl. 

At higher latitudes where the foreground reddening 
toward the target stars seldom exceeds 0.35 mag corresponding to the 
diffuse cloud regime, the present survey seldom finds CO emission 
except along sightlines at \EBVl\ $> 0.5$ mag.  This is similar to
what occurs along the line of sight toward \zoph\ (HD149757) 
where \WCO\ = 1.7 K \kms, \EBV\ = 0.32 mag and \EBVl\ = 0.55 mag, but 
the inherent liklihood of finding lines of sight with such 
\EBVl\ at high latitude is much higher in the survey sample here
than for randomly chosen lines of sight. 

\section{Tracing absorbing gas in emission?}

Although it was an original goal of this work to test whether the 
behaviour predicted for absorbing gas seen in emission could actually 
be observed, the profound contamination of the survey by background 
emission from associated material complicates or moots this issue.  
Given the built-in bias for the target stars to be situated in front 
of the associated material, the question becomes one of ascertaining whether 
the emission from foreground absorbing CO may even be reliably 
identified in arcminute-diameter mm-wave emission fan beams. 

The behavior predicted for emission from the foreground absorbing
gas takes two forms, as shown in Fig. 6 of \cite{Lis07CO}. Most 
importantly,  the integrated intensity should vary as 
\WCO\ = $\int\TB {\rm dv} \approx 1$ K \kms\ N(CO)/$10^{15}\pcc$ 
with no threshold in N(CO) (which varies as N(\HH)$^2$) and little
dependence on the ambient number density in the host gas.    
The proportionality between \WCO\  and N(CO) is a consequence of weak 
rotational excitation as originally shown by \cite{GolKwa74}.  This 
behavior was actually seen in CO emission toward the sample 
of extragalactic mm-wave background sources where all the gas is 
in the foreground \citep{LisLuc98}.

Figure 5 at bottom shows the comparison between emitted \WCO\ and 
absorbing N(CO) together with some model results which serve to sketch 
the expected emission locii for the absorbing gas.  CO emission was 
detected along every line of sight, save one, for which 
N(CO) $\ga 4\times10^{14}\pcc$, corresponding approximately to 
an expected emission contribution \WCO\ $\ga 0.4$ K \kms.  
In Sect. 3.2 we estimated a rough {\it a priori} $3\sigma$ sensitivity 
limit of 0.22 K, making it appear that CO emission was actually 
detected to the extent expected, but the emission
is contaminated by an uncertain contribution from background gas.

The expectation for the diffuse foreground gas is that the datapoints 
should lie on or slightly below the model results which are shown 
(see Fig. 6 of \cite{Lis07CO}) and there is a component of the data,
about one-third of the detections, for which this is the case.  
It is these lines of sight for which the contribution from the
foreground gas has most likely, not definitely, been 
isolated.   However, many more  lines of sight with 
detected CO are highly overluminous in the ratio \WCO/N(CO).  
Given that this ratio is actually maximized in diffuse gas --
compare with \WCO/N(CO) = 500 K \kms\ $/ 3\times10^{19}\pcc \approx
15 $ K \kms $/10^{18}\pcc \approx 0.02$ K \kms/$10^{15}\pcc$ toward
Orion A or a typical dark cloud, respectively -- contamination by 
background gas is the only possible explanation for such behaviour.  

For the absorbing foreground gas, the CO-\HH\ conversion factor 
\WCO/N(\HH) should be small for lower N(\HH), only approaching
values \WCO/N(\HH)  = 1 K \kms/ $2\times10^{20}\pcc$  for
N(\HH) $\ga 5\times10^{20}\pcc$/(K \kms).  Fig. 5 at top shows 
the observed integrated intensity plotted against the absorbing 
column density of \HH.  The contribution from any underluminous 
diffuse foreground gas is presumably represented by the 
non-detections at N(\HH) $ \la 5\times10^{20}\pcc$
but there is not much of a margin of detectability.  
What is true is that essentially every line 
of sight with detected CO emission has \WCO/N(\HH) $ \ga 
1$ K \kms\ /$1-2\times10^{20}\pcc$ . 
Perversely, most of the very strongest CO emission is found 
along lines of sight with very small absorbing N(\HH) and N(CO), a sure
sign of contamination by background gas. The sole line of
sight with appreciable N(\HH) and no CO emission, HD217035, 
has log N(\HH) =  20.95 and log N(CO) = 14.57, a very small 
fractional CO abundance, and an inferred a foreground emission
contribution \WCO\ $<$ 0.4 K \kms.

\begin{table*}
\caption[]{Statistics of \EBV, \EBVl\ and \WCO$^1$ }
{
\begin{tabular}{ccccccccccc}
\hline
  &&&&&&&\multicolumn{2}{c}{CO undetected} &\multicolumn{2}{c}{CO detected} \\
$|{\rm b}|_<$ & N  & \EBVl $>$ 1 mag &  $<$\EBVl$>/<{\rm l}>$ & CO? 
  &  $<$\WCO$>/<{\rm l}>$ & a \EBVl/2\WCO$^2$ & $<$\EBV$>$ & $<$\EBVl$>$ & $<$\EBV$>$ & $<$\EBVl$>$ \\
 & \# & \#  & mag kpc$^{-1}$ & \# & K \kms\ kpc$^{-1}$ & H $\pcc$ (K \kms)$^{-1}$ &mag&mag&mag&mag  \\
\hline
 3\degr& 71& 21& 1.4& 34& 14.1 & $2.8\times10^{20}$& 0.21& 0.48& 0.48& 1.87 \\
 5\degr& 63& 16& 1.66& 29& 21.3& $2.3\times10^{20}$& 0.19& 0.41& 0.45& 1.98 \\
10\degr0& 49& 14& 2.4& 21& 35.7& $1.9\times10^{20}$& 0.16& 0.38& 0.41& 2.46 \\
15\degr& 36& 11& 2.8& 16& 38.6 & $2.1\times10^{20}$& 0.13& 0.42& 0.39& 2.76 \\
20\degr& 16&  1& 1.5&  4& 30.0 & $1.5\times10^{20}$& 0.12& 0.28& 0.46& 0.58 \\
\hline 
\end{tabular}}
\\
$^1$For explanation of table entries see Sect. 4.2 and 4.3 \\
$^2$a $= 5.8 \times 10^{21} ~{\rm H} \pcc ~{\rm mag}^{-1}$ \\
\end{table*}



\begin{figure}
\psfig{figure=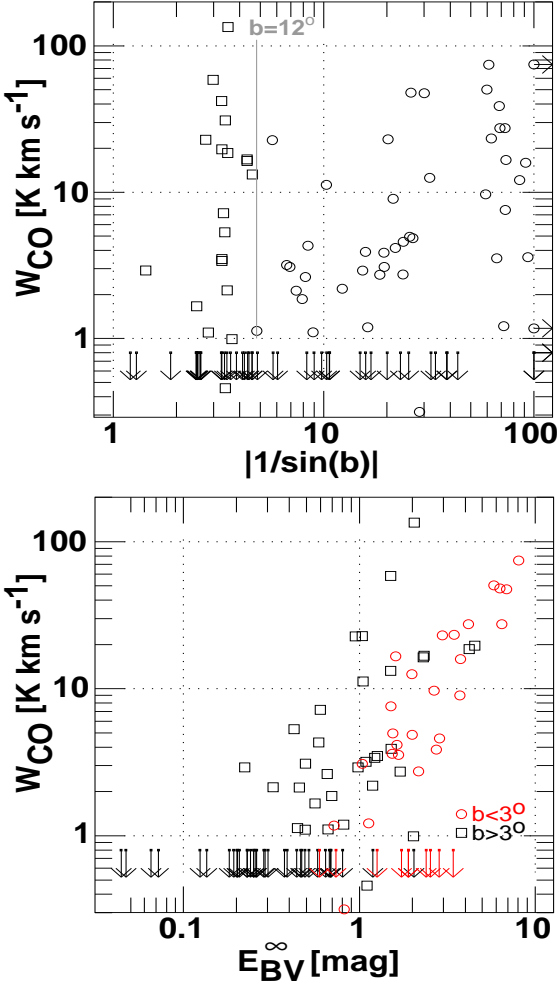,height=13.2cm}
\caption[]{Variation of the integrated intensity of the CO emission \WCO. 
Top: $vs.$ the cosecant of the galactic latitude.  Bottom: $vs.$
the reddening out to infinity from \cite{SchFin+98} where lines
of sight at lower latitudes are shown as red open diamonds.  Lines
of sight lacking detectable emission are indicated symbolically.}
\end{figure}

\begin{figure}
\psfig{figure=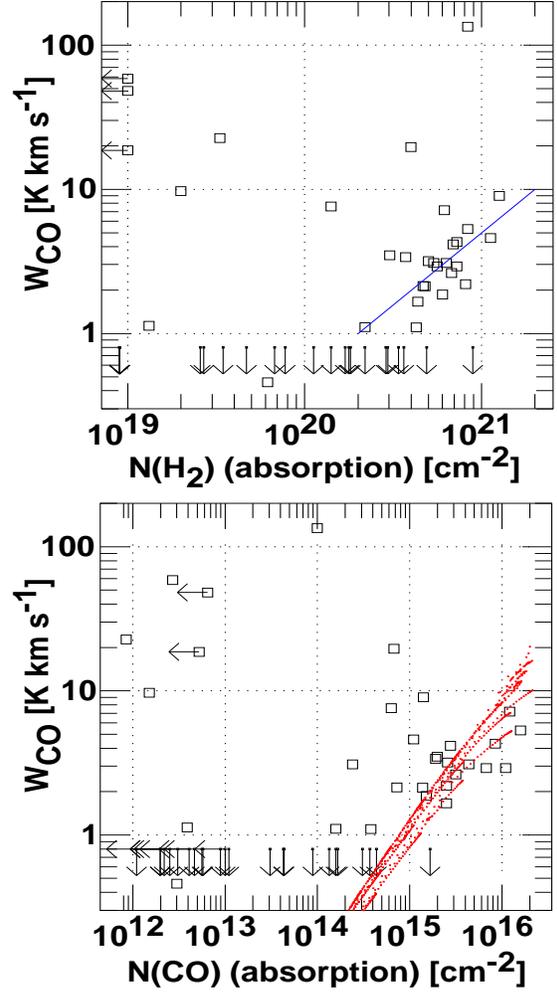,height=13.2cm}
\caption[]{Variation of the integrated intensity of the CO emission \WCO. 
 Top: {\it vs.}  the column density of \HH\ seen in {\it absorption}. 
The straight grayed (blue) line represents a CO-\HH\ conversion factor
\WCO/N(\HH)  $ = 2\times10^{20}\pcc $/(K \kms).   Bottom: \WCO\ plotted against
the column density of CO seen in {\it absorption}.  The dotted (red) locii
at bottom correspond to the model results for n(H) = 128 and 256 $\pccc$
shown in Fig. 6 of \cite{Lis07CO} and indicate the expected behaviour for
foreground gas observed in emission.   Lines of sight lacking detectable 
emission are indicated symbolically.}
\end{figure}

\section{Summary}

To test the predictions made by absorption line observations, and to provide
complementary information such as very highly velocity-resolved line
profiles, we conducted a survey of the mm-wave J=1-0 \cotw\ emission toward 
110 stars
which have in the past served as prime targets for optical/{\it uv}
absorption-line studies of the interstellar molecules \HH, CO, CH, CH\p\ and 
\H3\p.  The sample of included sightlines was constructed on the bases of
inclusiveness and practicality, {\it i.e.} $\delta > -25$\degr.  As discussed 
in Sect. 3.1,  all of the profiles with detected interstellar emission are 
online in machine readable form  along with an additional fourteen profiles
of \coth.  Thumbnail sketches of the spectra are shown in Figs. 1-3 and the
observations are summarized in Tables 1-4.

The CO emission results were compared with other line of sight parameters such
as the total reddening out to infinity, \EBVl, and the absorbing column 
densities of \HH\ and CO, see Sect. 4 and 5.  In doing so we showed that the 
survey sightlines generally are biased toward very high mean density of 
material (total reddening per unit length typically 2 - 5 times the local
galactic average, see Table 5 and Sect. 4) and comparably large amounts
of CO emission .  

This can be understood if
the targets are physically associated with the neutral material causing 
the over-density,  presumably because they are of early spectral type and 
were further selected because they returned interesting spectra.  In this
case the target stars must in general also be interacting with and influencing 
the associated material, which to this extent can not be entirely 
typical of the interstellar gas as a whole.  The enormous 
scatter in N(CO) at fixed N(\HH) which is typical of {\it uv} absorption 
line surveys may be exaggerated owing to the photoionizing radiation 
of the target stars.  This could also explain why selective photodissociation 
and fractionation of \coth\ are observed in somewhat different degrees in 
absorption line surveys toward stars and toward extragalactic continuum sources 
(although small-number statistics may also play a role).

Because the target stars have also been selected to obey the diffuse cloud 
limit \AV\ $\la $ 1 mag, any overendowment of associated neutral material must
occur preferentially behind them.  This general scenario is manifested in 
the detection of some very strong (20-40K)
CO lines typical of natal giant molecular clouds near H II regions, and some
heavily saturated lines typical of dark clouds, behind any of which the target
stars would not have been suitable candidates for optical/{\it uv} absorption
line studies in the first place.   { Lines of sight with detectable
CO emission are somewhat more highly reddened in the foreground of the target
star (0.25 mag) but much more heavily reddened behind (1.5 mag). 
The effect of association and preferential foreground placement creates 
complications for observations of CO in emission which lack an innate 
ability to discriminate between foreground and background material in 
the vicinity of the star.

One of the original goals of this survey was to compare direct observations with the 
prediction, based on the observable properties of absorption line gas, that 
\WCO\ $\approx 1$ K \ kms\ N(CO)$/10^{15} \pcc$, with little dependence
on the number density of the host gas.  The physical origin of such a 
proportionality resides in very general properties of the weak-excitation limit 
but the normalization depends on the particular properties of the diffuse gas.  
For instance, \WCO/N(CO) is some 50-100 times higher in diffuse gas 
than when the CO rotation ladder 
is thermalized, whether in dark clouds (\WCO\ = 15 K \kms, N(CO) = $10^{18}\pcc$)
or toward Orion A (\WCO\ = 500 K \kms, N(CO) = $3\times 10^{19}\pcc$).

However it seems more likely that the ratio of \WCO\ observed in emission to 
N(CO) seen in absorption is best used as a gauge of whether the detected 
emission can reasonably be associated with the foreground absorbing material.  
Along many lines of sight surveyed here, which have \WCO/N(CO) much higher
than even the already very large value which applies to diffuse gas (see Fig. 5) 
this is manifestly not so, owing to background contamination.  In these cases, 
the observed CO emission contains information on the environent of the star, 
but not necessarily on the foreground gas column.  Along other lines of sight,
the observed emission may contain a substantial though not necessarily 
exclusive or even dominant contribution from foreground material.

When the foreground gas can be isolated, CO emission offers the prospect
of very highly velocity-resolved line profiles, and, perhaps uniquely,
the opportunity to image the host body of the absorbing gas column with
high spatial and spectral resolution.  
These are laudable goals, but, as we have shown here, great care must be 
exercized in deciding whether they are actually achievable in any 
individual case.  We are currently trying to find sightlines toward
bright stars which are good candidates for mapping the foreground gas.
In the meantime, a more fruitful approach to mapping diffuse clouds
in CO might be to concentrate on those which occult extragalactic
background sources, for which confusion between foreground and background
material is not an issue.

\appendix{}
\section{Telluric emission}

\begin{table*}
\caption[]{Sightlines lacking interstellar CO J=1-0 emission$^1$}
{
\begin{tabular}{lccccccccc}
\hline
Target & l & b & \EBVl & \EBV &v$_{\rm tel}$ & rms & V$_{\rm lsr}$ - V$_{\rm hel}$ 
 & N(CO) & N(\HH)  \\
HD &$\deg$ &$\deg$ & mag & mag & \kms\ & K & \kms\ & $\pcc$ & $\pcc$  \\
\hline
13994&134.39 & -3.42 &0.64&0.62&-6.17 &0.098   &3.31&&  \\
21856&196.46 &-50.96 &0.07&0.19&-29.36 &0.124   &-14.63&& 20.04 \\
22951&158.92 &-16.70 &0.46&0.24&-14.08 &0.087  &-5.96&14.22 &20.46  \\
23408&166.17 &-23.51 &0.69&0.02&-18.11 &0.098   &-8.66&& 19.75 \\
23480&166.57 &-23.75 &0.69&0.08&-17.98 &0.169   &-8.79&&20.12  \\
23630&166.67 &-23.46 &0.23&0.00&-18.32 &0.088&-8.79& $<$12.34 &19.54    \\
23850&167.01 &-23.23 &0.12&0.01&-18.56 &0.162   &-8.87&&  \\
23840&159.46 &-15.03 &0.21&&-33.41 &0.055    &-5.94&& \\
24912&160.37 &-13.11 &0.29&0.33&-12.78 &0.066 &-6.01&13.49 &20.53   \\
30614&144.07 & 14.04 &0.23&0.30&-14.38 &0.073  &2.51&14.49 &20.34  \\
34078&172.08 & -2.26 &2.40&0.52&-6.72 &0.126    &-8.21&& \\
35149&199.16 &-17.86 &0.14&0.11&-14.71 &0.107&-16.12& $<$13.00 &18.30    \\
36486&203.86 &-17.74 &0.24&0.07&-14.17 &0.116 &-16.93& $<$12.04 &14.74  \\
37128&205.21 &-17.24 &0.30&0.08&-14.13 &0.085 &-17.12& $<$12.30 &16.28   \\
42087&187.75 &  1.77 &0.59&0.36&-30.60 &0.098   &-11.82&&  \\
57060&237.82 & -5.37 &0.52&0.18&-5.79 &0.157  &-18.84& $<$12.67 &15.78  \\
57061&238.18 & -5.54 &0.58&0.15&-5.89 &0.158  &-18.85& $<$12.61 &15.48  \\
58510&235.52 & -2.47 &0.74&&-23.78 &0.138   &-18.53&&  \\
121968&333.97 & 55.84 &0.07&0.07&29.95 &0.145  &7.56& $<$12.30 &18.70 \\
143275&350.10 & 22.49 &0.25&0.16&18.28 &0.099  &9.65&12.49 &19.41  \\
144217&353.19 & 23.60 &0.27&0.20&35.21 &0.116 &10.54&13.63 &19.83  \\
144470&352.75 & 22.77 &0.39&0.22&34.93 &0.120  &10.38&12.95 &20.05 \\
145502&354.61 & 22.70 &0.49&0.27&34.88 &0.120  &10.86&12.76 &19.89  \\
164353& 29.73 & 12.63 &0.20&0.12&10.47 &0.086  &17.54&13.04 &20.26  \\
166937& 10.00 & -1.60 &3.12&0.24&24.39 &0.131   &12.37&&  \\
167971& 18.25 &  1.68 &3.42&1.08&5.93 &0.129   &14.56&&  \\
170074& 75.06 & 23.71 &0.05&&14.80 &0.090   &18.81&&  \\
177989& 17.81 &-11.88 &0.20&0.23&-1.67 &0.123 &12.41&14.64 &20.23  \\
181615& 21.84 &-13.77 &0.18&&-2.31 &0.153   &12.81&&  \\
184915& 31.77 &-13.29 &0.26&&-2.70 &0.101   &14.41&& 20.31 \\
192639& 74.90 &  1.48 &1.90&0.66&2.07 &0.131 &17.46&14.13 &20.69    \\
193237& 75.83 &  1.32 &1.73&0.63&2.23 &0.102    &17.34&& \\
197592& 30.70 &-32.15 &0.04&&-13.21 &0.121   &9.93&&  \\
198478& 85.75 &  1.49 &2.85&0.54&1.09 &0.102   &16.06&&  \\
198781& 99.94 & 12.61 &0.47&0.35&6.23 &0.066&14.51&15.22 &20.56   \\
199579& 85.70 & -0.30 &2.54&0.37&0.23 &0.112 &15.84&& 20.36  \\
201345& 78.44 & -9.54 &0.19&0.32&-4.69 &0.064 &15.41& $<$12.40 &19.43  \\
203064& 87.61 & -3.84 &1.19&0.28&-1.71 &0.092   &15.03&& 20.29 \\
206773& 99.80 &  3.62 &2.04&0.50&0.55 &0.101 &13.67&14.20 &20.47  \\
209339&104.58 &  5.87 &0.68&0.37&1.96 &0.105  &12.83&13.95 &20.25   \\
209975&104.87 &  5.39 &0.80&0.38&1.80 &0.080   &12.71&& 20.08 \\
217035&110.25 &  2.86 &1.18&0.76&-0.26 &0.090 &11.09&14.57 &20.95  \\
217675&102.21 &-16.10 &0.25&0.05&-10.33 &0.135 &10.11&12.75 &19.67   \\
218915&108.06 & -6.89 &0.29&0.29&-5.68 &0.134 &10.30&13.64 &20.15  \\
224572&115.55 & -6.36 &0.38&0.17&-5.99 &0.113   &8.44&&  \\
\hline
\end{tabular}}
\\
$^1$For explanation of table entries see Sect. 3 \\
All column densities are logarithmic. \\
\end{table*}

\begin{figure}
\psfig{figure=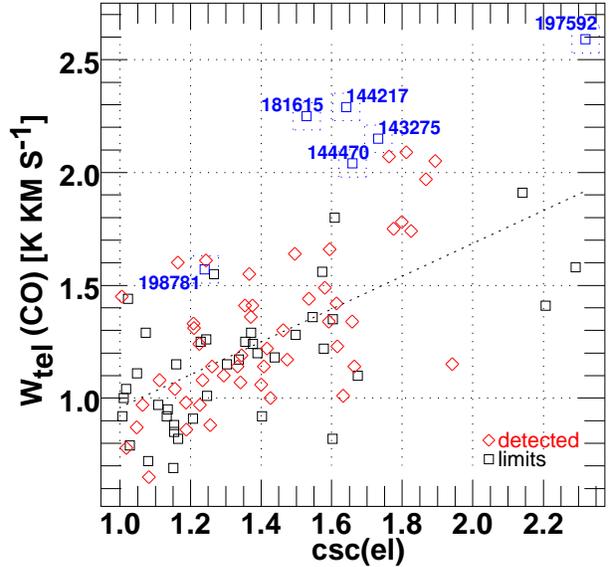,height=7.6cm}
\caption[]{Integrated intensity of the telluric CO emission $vs.$ the 
cosecant of the elevation.  Lines of sight with detected and undetected
interstellar emission are shown as open rectangles and (red) diamonds,
respectively.  A few lines of sight from Table 1 without obvious 
interstellar emission are distinguished by relatively strong telluric 
emission and are labelled (in blue).}
\end{figure}

\begin{figure}
\psfig{figure=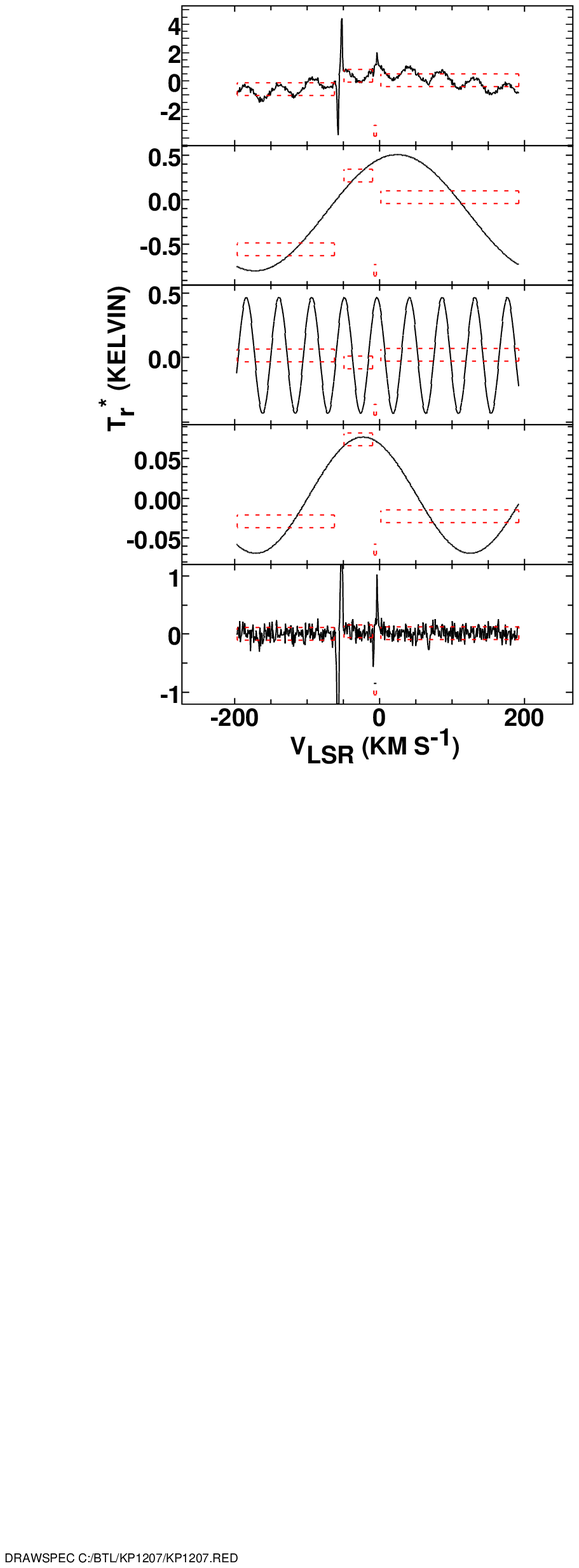,height=13cm}
\caption[]{Decomposition of an instrumental baseline.  Top: observed
spectrum toward HD2905, after co-adding both polarizations.  Middle upper
to lower: 
instrumental sine-wave baselines of period 405, 35.4 and 293 \kms,
found and removed in that order.   Bottom: spectrum after baseline 
subtraction.  The dashed, inset (red) rectangles show the regions 
of the spectrum used for baseline detection.  The spectra have all 
been boxcar smoothed over 8 channels for purposes of illustration.}
\end{figure}

A narrow (FWHM of 300 kHz or 0.8 \kms) telluric CO emission line
appears with \WCO\ = 1-2 K in all frequency-switched spectra.  In
some spectra, especially at low galactic latitude (for instance
toward HD169454), the telluric emission is inseparable
from that originating in the Galaxy.  In principle,
this could conceal an interstellar line if observations are not made
at two epochs widely spaced during the year.  In practice the telluric
line may appear at velocities which are much larger than those which
are likely to appear except at very low galactic latitude.  

The velocity at which the telluric line appears in our spectra is
given in Tables 1-3 and the measured integral of the telluric 
emission is shown in Fig. A.1, where lines of sight with and without
detected interstellar emission are separately noted.  Scatter in the
integrated intensity is far larger than that attributable to the
typical rms noise level which (see Sect. 2) provides signal/noise
levels of about 20:1 with a single channel rms noise level of 0.1 K.
The telluric emission fills the entire forward response of the telescope
so much of the scatter in Fig. A.1 must be attributed to random errors 
of calibration.  A
few lines of sight with non-detections of interstellar emission and
exceptionally large telluric CO emission are noted in Fig. A.1.
It seems likely that no substantial amount of interstellar emission
was missed during the survey.

Note that the present survey includes non-detections in
directions where  interstellar lines were claimed in the Figures
of \cite{KnaJur76}, for instance toward P Cyg (HD193237) and 55 Cyg (HD198478).
In these cases, and for the most negative-velocity component 
shown by them toward \zoph\ and the highest-velocity components shown
toward 9 Cep (HD206165) and HD207538, the claimed interstellar line 
is actually telluric in origin.  Other telluric lines are included in
their Table of claimed detections.  This may be verified by noting the dates 
on which the data of \cite{KnaJur76} were taken.  

\section{Baseline subtraction}

Figure A.2 shows the spectrum observed toward HD2905, along with the 
three sine-wave components which comprise the spectral baseline.  
These were detected by least-squares fitting and subtracted in order 
from uppermost downward.  Only
the middle component with period near 35 \kms\ is capable of 
concealing a typical isolated interstellar component in spectra
away from the galactic equator.

\section{Unfolding}

Surveys like that done here are impractical if a clean 
reference sky position must be found anew for each source (or for too many 
sources), making use of frequency-switching obligatory.  Figure A.3 
shows two examples of unfolding frequency-switched spectra 
using the variant of the EKH dual-beam switching technique \citep{EKH} 
described in \cite{Lis97FS} and employed here.  The frequency 
switching interval in both
spectra is $\pm 1$ MHz, or a total of 5.2 \kms.  In both cases the
emission continuously spans substantially more than 5.2 \kms.

The advantages of this method of
observing, which allows faithful reconstruction of signals which are
broader than the switching interval, are that it obviates the need for a 
signal-free off-source reference switching position while permitting 
narrower frequency switching intervals when searching blindly
for the prospective emission profile.  Broadening these intervals
to avoid switching within a (known or unknown) signal may cause 
deterioration of the baseline to such an extent that detection of
even narrow lines is hindered.

The most obvious disadvantage of the EKHL reconstruction is that the 
${\sqrt 2}$ gain in signal/noise which accrues to the usual shift and 
add method of unfolding frequency-switched spectra is not fully 
realized.  Furthermore, unless the spectral baseline can be removed, 
reconstruction is prone to defects attributable to aliasing.

\begin{figure*}
\psfig{figure=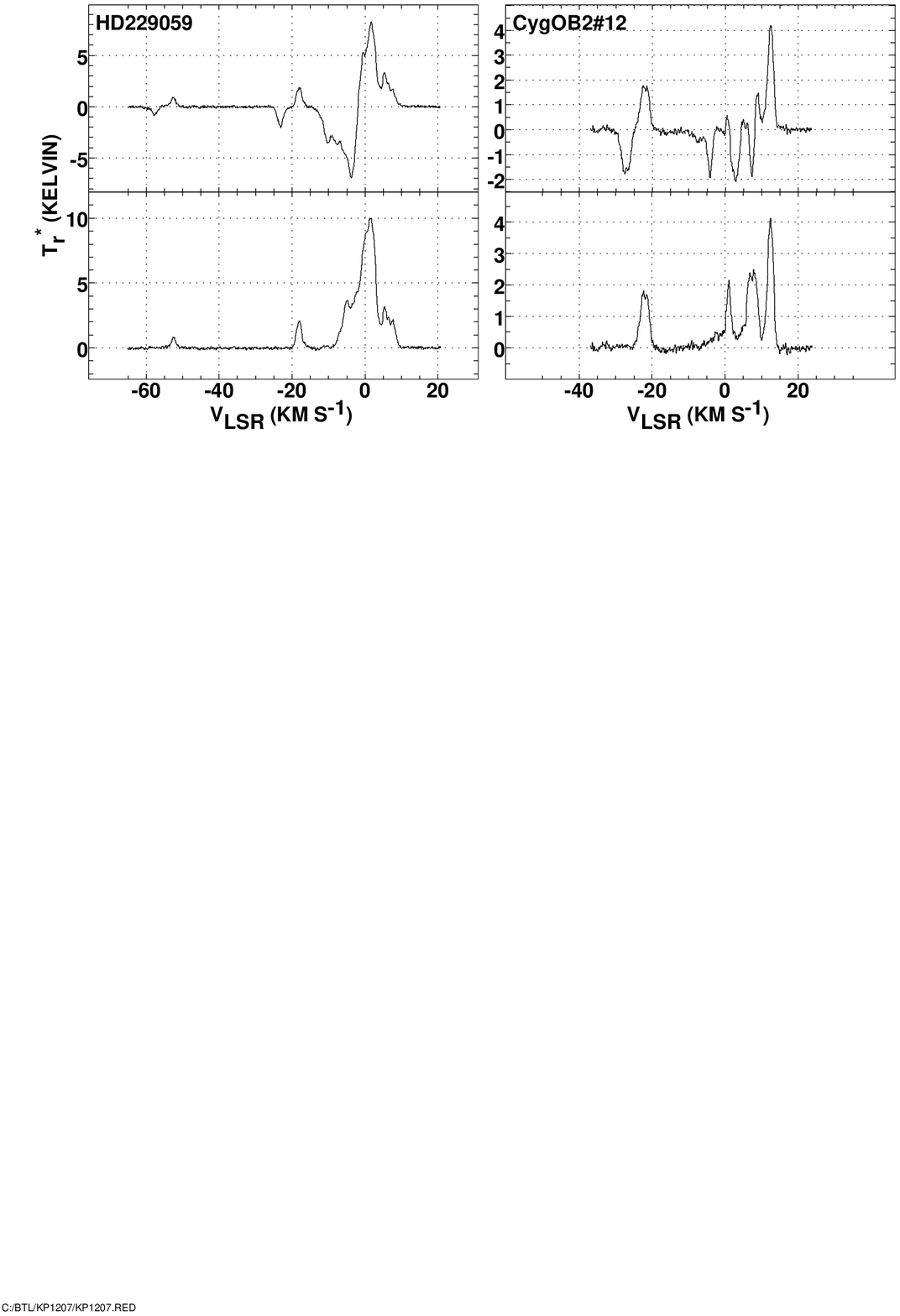,height=9cm}
\caption[]{Examples of baseline-subtracted frequency-switched (top) and 
unfolded spectra reconstructed with the EKHL algorithm described in 
\cite{Lis97FS}, toward two sources.  Compare the spectra at right with
Fig. 5 of \cite{McCHin+02}.}
\end{figure*}

Because frequency-switching of any kind is a convolution, some 
spectral components of the signal are unavoidably downweighted
corresponding to the smaller Fourier components of the convolution 
pattern which mathematically represents the shift and subtract 
process implemented in hardware.  With sufficient signal/noise 
and flat baselines, the EKHL reconstruction will recover
all components of the signal which are not entirely filtered out.  
However, if too much of the sky signal occurs with spectral 
frequencies which are heavily downweighted during the 
frequency-switching, reconstruction may be impeded by the
actual noise and baseline imperfections.  In our experience, such 
effects are not subtle after reconstruction and the only cure 
for them is to reobserve with a different frequency switch 
interval.  Only two lines of sight were affected
in such a way in this work.

\begin{acknowledgements}

The National Radio Astronomy Observatory is operated by Associated 
Universites, Inc. under a cooperative agreement with the US National 
Science Foundation.
The Kitt Peak 12-m millimetre wave telescope is operated by the
Arizona Radio Observatory (ARO), Steward Observatory, University
of Arizona.  I am grateful to the ARO Director, Dr. Lucy Ziurys. 
for awarding the observing time necessary to perform these 
observations and to the ARO staff and 12m operators who keep 
the telescope running at such a laudably high level.  This work 
profited from many discussions with Jerome Pety and Robert Lucas and
remarks by the referee provided an impetus to improve the
manuscript.

\end{acknowledgements}
 

\end{document}